\documentclass{ws-procs975x65}

\usepackage{wrapfig}
\usepackage{graphicx}
\usepackage{amsmath}
\usepackage{bm}
\usepackage{amsfonts}

\newcommand{\be}{\begin{equation}}
\newcommand{\ee}{\end{equation}}
\newcommand{\bes}{\begin{equation*}}
\newcommand{\ees}{\end{equation*}}
\newcommand{\bea}{\begin{eqnarray}}
\newcommand{\eea}{\end{eqnarray}}
\let\non\nonumber

\newcommand{\CG}{{\cal G}}

\newcommand{\CL}{{\cal L}}

\newcommand{\CO}{{\cal O}}

\newcommand{\IR}{{\mathbb R}}

\newcommand{\Imm}{\mbox{Im} \,}
\newcommand{\open}{\left( \begin{array}{cc} }
\newcommand{\opens}{\left( \begin{array}{cccc} }
\newcommand{\close}{\end{array}\right)}
\def\vev#1{\langle{#1}\rangle}

\newcommand{\bca}{\begin{cases}}
\newcommand{\eca}{\end{cases}}

\newcommand{\half}{\frac{1}{2}}

\begin{document}

\title{Holographic Fermi surfaces and bulk dipole couplings}

\author{David Guarrera and John McGreevy}

\address{Center for Theoretical Physics \\ Massachusetts Institute of Technology \\ Cambridge, MA  02139 USA}
\date{MIT-CTP/4205}


\begin{abstract}

Non-Fermi liquids can be studied using holographic duality.
The low energy physics of a holographic Fermi surface is controlled by an emergent scale invariance.
After reviewing these developments, we 
generalize the holographic calculation 
to include in the bulk action
the leading irrelevant operator, which is a dipole coupling 
between the spinor field and the background gauge field.
We find that this dipole coupling changes the attainable 
low-energy scaling dimensions, and 
changes the locations of the Fermi surfaces in momentum space.
The structure of the holographic framework for non-Fermi liquids
is, however, robust under this deformation.

\end{abstract}

\bodymatter

\section{Introduction}

Holographic duality is a wonderful development arising from string theory
which offers a new perspective on strongly coupled quantum systems.
The duality solves certain such systems in terms of an auxiliary theory
of gravity in one extra dimension.  This extra dimension plays the role of
the renormalization group scale, and
Einstein's equation is an RG evolution equation\footnote
{For reviews of these developments in the spirit of the present work, see 
\citelow{Hartnoll:2009sz , McGreevy:2009xe, 2009arXiv0909.3553H}.}.

The question ``To which strongly coupled systems does the duality apply?"
has not been settled.  
The best understanding at the moment is for certain relativistic, supersymmetric gauge theories.  
Though these systems are not yet known to be realized 
in Nature, 
this shortcoming can be mitigated
by judiciously chosen questions about long-distance physics,
where the same universal behavior can arise from systems which are 
very different microscopically.

The basic dictionary between the `boundary' quantum system and
the `bulk' gravity theory is as follows.  
The gravity theory is classical when the boundary system
has a large number (which we will call $N^2$) of degrees of freedom per spatial point.
There is a one-to-one correspondence between local operators
in the boundary theory 
and fields in the bulk.
Correlation functions of these local operators may be computed
by solving classical wave equations for the corresponding bulk field.
To study thermodynamic equilibrium of the boundary system, one 
places a static black hole in the bulk geometry.

Attempts have been made to apply this technology, with some success, to the
quark-gluon plasma, quantum-critical transport \cite{Herzog:2007ij, Hartnoll:2007ih},
and ultracold atoms at unitarity \cite{Balasubramanian:2008dm, Son:2008ye, Adams:2008wt, Maldacena:2008wh, Herzog:2008wg}.

One is led to wonder whether this apparatus can be 
applied to the study of metallic states.
Most of the present understanding of metals relies on Landau's Fermi Liquid 
ansatz: one supposes that the low-excitations are long-lived electron-like quasiparticles
near the Fermi surface in momentum space.  This ansatz results 
in a powerful and robust low-energy effective field theory with many successes,
but there are systems to which it does not apply.
The feature of such ``non-Fermi liquids'' on which we focus
is the presence of a sharp Fermi surface of gapless excitations,
which lack a description in terms of long-lived quasiparticles.


In 
\citelow{Lee:2008xf, Liu:2009dm, Cubrovic:2009ye , Faulkner:2009wj, Faulkner:2010da}, 
it was shown that holographic duality can describe Fermi surfaces.
Indeed, they arise from the minimal ingredients necessary to pose the problem.
The system under study is, microscopically, a 2+1-dimensional relativistic conformal field theory (CFT);
this means that the dual geometry is asymptotically four-dimensional 
anti-de Sitter space, $AdS_4$.
Assuming that this CFT has a conserved $U(1)$ current (a proxy for fermion number), 
the gravity theory must include a massless photon field $A_\mu$.
In order to study fermion response, the CFT must contain a fermionic operator,
with some charge $q$ under the $U(1)$ current, and some scaling dimension $\Delta$
at the short-distance (UV) fixed point.  
To introduce a finite density of $U(1)$ charge,
one studies the charged black hole in $AdS_4$. 
This finite density breaks the Lorentz symmetry and scaling symmetry of the boundary theory. 
The zero temperature groundstate is described by an extremal black hole.

In Ref.~\citelow{Faulkner:2009wj} it was shown that this 
finite-density ground state exhibits an emergent scaling symmetry which 
is manifested by the fact that the near-horizon region of the extremal black hole geometry
is 
$AdS_2 \times \IR^2$.  
By holographic duality, this geometry is dual to a fixed point field theory, which we call the {\it IR CFT}.
The IR scale transformations act on time but not space, and in this sense the IR CFT has 
dynamical exponent $z=\infty$.
This emergent quantum critical behavior
governs the low-frequency fermion response.
The retarded fermion Green's function
exhibits a Fermi surface, near which it
takes the form\be
 \left< \psi^\dagger(k) \psi(-k) \right>=  \frac{h_1}{k_\perp- \frac{1}{v_f} \omega -h_2 e^{i \gamma} \omega^{2 \nu}}
\ee
with $k_\perp \equiv |\vec k|-k_f$, and real constants $h_1, v_f, \gamma, \nu$. 
$\nu$ is a scaling dimension in
the IR CFT. In this way, the emergent conformal symmetry controls the dispersion of these excitations. This duality construction provides a large theoretical playground of non-Fermi liquid fixed points, 
giving a handle on a difficult strong coupling problem. In particular, the case with $\nu=1/2$ has several features in common with the  ``strange metal" phase of high $T_c$ superconductors.

In this work\footnote{
 {\bf Note Added 1:} After this paper was completed, but several months before it appeared on the arXiv, 
the papers \citelow{Edalati:2010ww, Edalati:2010ge} appeared,
which are the first published papers to study the effects of the magnetic dipole coupling $g_m$.
We feel that our perspective on the subject is still worth sharing.
The papers \citelow{Edalati:2010ww, Edalati:2010ge} do not study $g_e$.
}\footnote{We restrict our attention to 2+1-dimensional field theories,
in which case the gravity theory lives in four dimensions.}, we consider 
effects on these holographic Fermi surfaces 
of an additional dimension five operator
 \be
 \bar{\psi} (g_m + g_e \Gamma) \Gamma^{M N} \psi F_{MN}
 \ee 
 in the bulk action
 corresponding to magnetic and electric dipole moments for the bulk fermions
 ($\Gamma \equiv \Gamma^{\underline{t}} \Gamma^{\underline{r}} \Gamma^{\underline{x}} \Gamma^{\underline{y}}, 
 \Gamma^{M N} \equiv \half [\Gamma^M, \Gamma^N]
 $). 
 Our motivation for this addition is several-fold. First, we would like to investigate the robustness of the previous discoveries. 
 The choice of action for the bulk fields in the calculations described above
was motivated by Landau-Ginzburg-Wilson Naturalness: 
 the lowest-dimension operators respecting the symmetries were used.  
 This Naturalness criterion is usually enforced by the renormalization group.
In the bulk gravity theory, the status of this principle is not clear
because of our poor understanding of quantum gravity.
This motivates an exploration of the sensitivity of the results to 
RG-irrelevant bulk couplings.

In particular, our previous calculations used  the canonical Dirac action.
The two-point function is insensitive to self-interactions of the spinor field
at leading order in the $1/N$ expansion.
However, there exists a large class of higher-dimension operators 
which are quadratic in the spinor
that can possibly change these conclusions. This work on dipole couplings is an attempt to 
investigate systematically the effects of such operators.  Do these higher dimension operators in the bulk drastically alter the existence of Fermi surfaces in the boundary? We will find that they do not, but rather the main effects of the dipole couplings are to change the IR $AdS_2$ scaling dimensions and to change the locations of Fermi surfaces in $k$ space (which we will find numerically)\footnote{{\bf Note Added 2:} However, \citelow{Edalati:2010ww, Edalati:2010ge} observe numerically that 
at larger values of $g_m$ than are studied here, 
Fermi surfaces are not present, and further that the low frequency spectral weight is suppressed.  
We make some comments on these points at the end of this paper.}.
As such, we have constructed a much larger parameter space of non-Fermi liquids for study.
 
 The dipole moment couplings are a natural starting point for an 
 exploration of irrelevant operators, since the structure of the calculation is largely preserved.
Further, they are generic in the following sense. The $AdS_4$ under study 
frequently arises as a factor in a higher-dimensional bulk spacetime $AdS_4 \times X$, where $X$ is compact. There will be an effective theory of the light modes (or a ``consistent truncation") on $AdS_4$. Even when one starts with the simplest Dirac Lagrangian on $AdS_4 \times X$, 
if the charge on the black hole comes from the Kaluza-Klein gauge field,
such dipole terms generically appear in the low energy effective theory on $AdS_4$  \footnote{We thank N. Iqbal for this point.}. 
 These couplings also appear in some string theory realizations of the 
 ingredients described above \cite{Ammon:2010pg,Bah:2010yt,Bah:2010cu,Liu:2010pq}.

\section{Fermion Green's Functions from Holography}
In the next three sections, we review previous work \cite{Liu:2009dm, Iqbal:2009fd, Faulkner:2009wj}
on the holographic computation of two-point functions of fermionic operators in a $2+1$ dimensional boundary CFT with a finite $U(1)$ charge density. For simplicity, we work at zero temperature. 

In the bulk, this ensemble corresponds to a black hole in $AdS_4$ charged under a $U(1)$ gauge field.
We employ the bulk action
\be
S_{bulk}= \frac{1}{2 \kappa^2} \int d^4 x \, \sqrt{-g} \left[ \mathcal{R} + \frac{6}{R^2} - \frac{R^2}{g_F^2} F_{MN} F^{MN} \right]
\ee
where $R$ is the AdS radius, $\kappa$ the Newton's constant and $g_F$ the gauge coupling. 
More specifically, the relevant solution is the charged $AdS_4$ black hole, 
\be
ds^2 = \frac{r^2}{R^2} (-f dt^2 + d \vec{x}^2) + \frac{R^2}{r^2} \frac{dr^2}{f} ,
~~~~  f = 1 + \frac{Q^2}{r^4} - \frac{M}{r^3} ,
~~~~ A_t = \mu (1- \frac{r_0}{r})
\ee
with $Q,M$ the black hole charge and mass respectively and $\mu \equiv g_F Q/(R^2 r_0^2)$. $r_0$ is the outer horizon, {\it i.e.}  the largest solution to $f(r_0)=0$. In the boundary, this geometry corresponds to a theory with finite charge density and temperature
\be
\rho = \frac{2 Q}{\kappa^2 R^2 g_F} , ~~~~
T= \frac{3 r_0}{4 \pi R^2} \left( 1 - \frac{Q^2}{3 r_0^4} \right)~~~.
\ee
At extremality, the inner and outer horizons merge into a double zero of $f$ and 
\be
M=4 \left( \frac{Q}{\sqrt{3}} \right)^{3/2}, \, \, \,  Q= \sqrt{3} r_0^2 \Longrightarrow T=0~~~.
\ee
We will work in units with $R=1$. In addition, in our numerical work, we will often put the horizon at $r_0=1$ and set $g_F=1$. 

We want to study the Dirac equation in the bulk,
\be
\Gamma^{M} D_M \Psi - m \Psi =0~~~.
 \ee
Here $\Gamma^M$ is related to the usual flat space gamma matrix by a factor of the vielbein, $\Gamma^M=e^M_{\phantom M \underline{\mu}} \Gamma^{\underline{\mu}}$
and 
$
 D_M  = \partial_M + {1 \over 4} \omega_{ab M} \Gamma^{ab} - i q A_M \
$
with $\omega_{ab M}$ the spin connection.
One can nicely cancel off the spin connection contributions to this equation by defining 
\be
\Psi = (-g g^{rr})^{-\frac{1}{4}} e^{-i \omega t+ i k_i x^i} \psi \label{fieldredefinition}
\ee
with $\vec{x}=(x,y)$ the spatial directions on the boundary. Substituting and rearranging, we get
\be
\sqrt{ \frac{g_{ii}}{g_{rr}} } ( \Gamma^{\underline{r}} \partial_r - m \sqrt{ g_{rr}}) \psi + i K_\mu \Gamma^{\underline{\mu}} \psi  =0,~~~~~~~K_\mu\equiv (-u, k_i)
\ee
with
\be
u \equiv \sqrt{ \frac{g_{ii}}{-g_{tt}}} \left( \omega + \mu_q (1-\frac{r_0}{r}) \right), ~~~~~~\mu_q \equiv \mu q.
\ee
This system of four coupled equations becomes simpler by rotating the $k$ momentum to be entirely in the $x$ direction 
(which we can do by rotational invariance) and by a choice of gamma matrices adapted to this frame, 
\be\label{eq:Diracbasis}
\Gamma^{\underline{r}}= 
\open -\sigma^3 & 0 \\ 0 & -\sigma^3 \close ~~  
\Gamma^{\underline{t}}= \open  i \sigma^1 & 0 \\  0 & i\sigma^1 \close ~~
 \Gamma^{\underline{x}}= \open -\sigma^2 & 0 \\ 0 & \sigma^2 \close  ~~
\Gamma^{\underline{y}}= \open 0 & -i \sigma^2  \\   i\sigma^2 & 0 \close ~~.
\ee
Defining $\psi=\left( \begin{array}{c} \Phi_1 \\ \Phi_2 \end{array} \right)$ and rearranging gives
\be
(\partial_r + m \sqrt{g_{rr}} \sigma^3 ) \Phi_\alpha= \sqrt{ \frac{g_{rr}}{-g_{tt}}} \left( \omega + \mu_q (1-\frac{r_0}{r}) \right)  i \sigma^2 \Phi_\alpha + \sqrt{ \frac{g_{rr}}{g_{ii}}} k (-1)^\alpha \sigma^1 \Phi_\alpha \label{secondDirac}
\ee
with $\alpha=1,2$. This gives two decoupled, real $2 \times 2$ equations. 

A solution of the Dirac equation, in the basis \eqref{eq:Diracbasis}, behaves near the boundary like
\be
\Phi_\alpha \sim a_\alpha r^m \left( \begin{array}{c} 0 \\ 1 \end{array} \right) + b_\alpha r^{-m} \left( \begin{array}{c} 1 \\ 0 \end{array} \right) \label{expansion}
\ee
In terms of the eigenspinors of $\Gamma^r$, $\psi_+, \psi_-$, we have
\be
\psi_+  \sim  A(k) r^m + \ldots  \, \, \, \,  , \, \, \, \,  A(k)= \left( \begin{array}{c} a_1 \\ a_2 \end{array} \right)  ~;~~~~
\psi_-  \sim D(k) r^{-m}+ \ldots  \, \, \, \,  , \, \, \, \,  D(k)= \left( \begin{array}{c} b_1 \\ b_2 \end{array} \right) ~.
\ee
These coefficients are related by a matrix $S$, 
\be
 \left( \begin{array}{c} b_1 \\ b_2 \end{array} \right) = 
 S \left( \begin{array}{c} a_1 \\ a_2 \end{array} \right) =
 \open s_1 & s_2 \\ s_3 & s_4 \close  \left( \begin{array}{c} a_1 \\ a_2 \end{array} \right) \label{matrixe} ~~.
 \ee
Since the two $\alpha$ equations are decoupled, we can choose independent boundary conditions that do not mix $\Phi_1$ and $\Phi_2$ giving $s_2=s_3=0$. The standard prescription for calculating the retarded Green's function gives
\be
G_R=i S \gamma^{t'}= - \open b_1/a_1 & 0 \\ 0 & b_2/a_2 \close \label{Green's}
\ee
with
\be
\Gamma^{\underline{t}'} = \open 0 & \gamma^{t'} \\ \gamma^{t'} & 0 \close = U \Gamma^t U^{-1}
\ee
where $U$ is the basis change between the $\Phi_\alpha$ basis and the chiral basis. 

For $0 \leq m < 1/2$, there exists an inequivalent holographic prescription 
using this same bulk action which describes a different boundary theory\cite{Klebanov:1999tb}. In this ``alternative quantization," the roles of the source ($a$) and response ($b$) are switched, and similar reasoning leads to $\tilde{G}_R^\alpha=-1/G_R^\alpha$. The boundary CFT in alternative quantization 
flows to the usual one upon adding $\CO^\dagger \CO$ to the CFT lagrangian, where $\CO$ is the operator dual to $\psi$.

\section{Low Frequency Limit of Retarded Green's Functions}

We are interested in frequencies small compared to the chemical potential, $\mu$. Naively, we should expand $\Phi_\alpha$ in a series in $\omega$. However, the term proportional to $\omega$ in (\ref{secondDirac}) is dominant at the horizon and so we cannot treat $\omega$ as a small perturbation there. To deal with this, split the $r$-axis into two regions, an inner region (with coordinate $\zeta$) and an outer region (with coordinate $r$). The inner region is specified by: 
\be
\label{defofzeta}
r-r_0 \equiv \frac{\omega R_2^2}{\zeta} \, \, \, , \, \, \, \epsilon < \zeta < \infty
\ee
and the outer is
\be
\frac{\omega R_2^2}{\epsilon} < r - r_0
\ee
with $R_2 \equiv 1/\sqrt{6}$ (recall that we have set the $AdS$ radius, $R=1$). The strategy now is to develop the solution as a perturbation series in $\omega$ with $\zeta$ in the inner region and $r$ in the outer region. Because the distinction between inner/outer involves $\omega$, the inner region equation no longer blows up in the $\omega \rightarrow 0$ limit and the perturbation series between the two regions is reshuffled.  The statement that results are independent of the matching point $\epsilon$ 
is a holographic version of the Callan-Symanzik equation.

Let us examine the lowest order solution in the inner region by taking the limit $\omega \rightarrow 0, \epsilon \rightarrow0, \omega R_2^2/\epsilon \rightarrow 0$. In this limit, the Dirac equation is
\be
(-\partial_\zeta + \frac{m R_2}{\zeta} \sigma^3) \Phi_\alpha=(\tilde \omega+  \frac{q e_3}{\zeta}) i \sigma^2 \Phi_\alpha+ \frac{R_2}{\zeta}  (-1)^\alpha \frac{k}{r_0} \sigma^1 \Phi_\alpha \label{adsDirac}
\ee
with $e_3 \equiv g_F/\sqrt{12}$. This is precisely the Dirac equation for a spinor in $AdS_2 \times \mathbb{R}^2$:
\be ds^2 = 
 {R_2^2 \over \zeta^2 } \left(- d \tau^2 +  d \zeta^2 \right) + {r_0^2 \over R^2} d\vec x^2 \ee
 in the presence of a constant electric field $e_3$.
Here $\psi_\alpha= (-g g^{\zeta \zeta})^{-1/4} \Phi_\alpha$, where $\zeta=0$ is the boundary, 
and the $AdS_2$ time coordinate is $\tau \equiv \lambda t$. 
To be consistent with the definition of $\zeta$
\eqref{defofzeta} the parameter $\lambda$ should be set equal to $\omega$;
the parameter $\lambda$ is introduced to avoid the awkwardness of rescaling 
the time coordinate by a frequency.
Relatedly, in \eqref{adsDirac} we have defined $\tilde \omega \equiv \omega/\lambda$; 
this is the frequency conjugate to the $AdS_2$ time coordinate.

We are interested in matching to the outer region where $\zeta \rightarrow 0$. Here, the equation becomes
\be
\zeta \partial_\zeta \Phi_\alpha = U_\alpha \Phi_\alpha
~~, ~~
U\equiv - \open -m R_2 & e_3 q+ (-1)^\alpha \frac{k R_2}{r_0} \\ -e_3 q+ (-1)^\alpha \frac{k R_2}{r_0}  & m R_2 \close ~.
\ee
This matrix has eigenvalues $\pm \nu_\alpha$ with
\be
\nu_\alpha=  \sqrt{\frac{k^2 R_2^2}{r_0^2}+m^2 R_2^2 - q^2 e_3^2}~. \label{firsteigen}
\ee
$\nu_\alpha$ is the scaling dimension in the IR CFT of 
the frequency-space operator dual to
$\psi_\alpha$. 
The corresponding eigenspinors are
\be
v_{\pm \alpha} = \left( \begin{array}{c} m R_2 \pm \nu_\alpha \\ e_3 q- (-1)^\alpha \frac{k R_2}{r_0} \end{array} \right)~~~.
\ee
At the boundary of $AdS_2$, the solution is therefore
\be
\label{ads2soln}
\Phi_\alpha^{I (0)} =   v_{- \alpha} \zeta^{-\nu_\alpha} + \CG_R^\alpha(\tilde \omega) v_{+ \alpha} \zeta^{\nu_\alpha} 
\ee
Generalizing (\ref{Green's}), the $AdS_2$ Green's function (in the presence of constant $E$ field) is $\CG_R^\alpha $ in \eqref{ads2soln}. 
To get the retarded function, we must set infalling boundary conditions at the horizon. 
The $AdS_2$ Dirac equation \eqref{adsDirac} is solvable\cite{Nabiltoappear}, and the associated retarded Green's function is 
\be
\CG_R^{\alpha}(\omega) =  e^{-i \pi \nu_\alpha} \frac{\Gamma(-2 \nu_\alpha) \Gamma(1 + \nu_\alpha- i q e_3)}{\Gamma(2 \nu_\alpha) \Gamma( 1 - \nu_\alpha- i q e_3) } \times \frac{(m + i  (-1)^\alpha \frac{k}{r_0} ) R_2 - i q e_3 -\nu_\alpha}{(m + i (-1)^\alpha \frac{k}{r_0} ) R_2- i q e_3 +\nu_\alpha} ( 2 
\tilde \omega) ^{2 \nu_\alpha} 
\ee
Note that by momentum conservation in $\mathbb{R}^2$, operators with different $k$ do not mix. 

Now we look at the outer region equations. Here, we can safely set $\omega=0$ to get the lowest order solution;  a basis of solutions is specified by the IR boundary condition
\be
 \eta_{\pm \alpha}^{(0)}
\sim v_{\pm \alpha} \left( \frac{R_2^2}{r-r_0} \right)^{\pm \nu_\alpha}~~.
\ee
Matching, we conclude
\be
\Phi_\alpha^{O (0)}=
 \eta_{+\alpha}^{(0)} + \CG_R^\alpha(\omega) \eta_{-\alpha}^{(0)} ~.
\label{thematching}
\ee
Now, in the outer region, we can perturbatively expand the linearly independent solutions
\be
\eta_{\pm \alpha}= \eta_{\pm \alpha}^{(0)} + \omega \eta_{\pm \alpha}^{(1)} + \ldots
\ee
where we have already solved for $\eta_{\pm \alpha}^{(0)} $. The higher orders can be obtained by solving the Dirac equation and requiring that the solution has no piece proportional to the lower order ones. Thus, the matching is entirely determined by the lowest order and we conclude
\be
\Phi_\alpha^{O}=\eta_{+ \alpha}+  \CG_R^\alpha(\omega) \eta_{- \alpha}
\ee
To know $\eta_{\pm \alpha}$ we must solve the Dirac equation everywhere -- we must have all the UV data. $\CG_R^\alpha(\omega)$ is however determined entirely by the IR CFT. At the boundary of $AdS_4$, we have (from (\ref{expansion}))
\be
\eta_{\pm \alpha}^{(n)}  \sim a^{(n)}  _{\pm \alpha} r^m \left( \begin{array}{c} 0 \\ 1 \end{array} \right) + b^{(n)}  _{ \pm \alpha}  r^{-m} \left( \begin{array}{c} 1 \\ 0 \end{array} \right) \label{allbdy}
\ee
giving the full Green's function perturbatively using (\ref{Green's})
 \be
 G_R^\alpha(\omega, k)=\frac{b_{+ \alpha} ^{(0)} + \omega b_{+ \alpha}^{(1)} + O (\omega^2)  +\CG_R^\alpha(\omega)(b_{- \alpha}^{(0)} + \omega b_{- \alpha}^{ (1)}  + O (\omega^2) )   }{ a_{+ \alpha}^{ (0)} + \omega a_{+ \alpha}^{ (1)}  + O (\omega^2)  +\CG_R^\alpha(\omega)(a_{- \alpha}^{ (0)} + \omega a_{- \alpha}^{(1)} + O (\omega^2) )  } \label{biggreen}
\ee
 

\section{Fermi Surfaces}
Let us suppose that there exist certain $k_f$ where $a_{+ \alpha}^{ (0)} (k_f)=0$. This will only happen for real $\nu_\alpha$. For small $k_\perp =k -k_f$ and small $\omega$, the Green's function (\ref{biggreen}) can be written
\be 
G_R^\alpha(\omega, k)\approx \frac{h_1}{k_\perp- \frac{1}{v_f} \omega -h_2 e^{i \gamma_{k_f}} 
\omega^{2 \nu_{k_f} }  } \label{fscorr}
\ee
with the (real) constants $h_{1,2}, v_f$ in \eqref{fscorr} determined by the UV data $a_\pm^{(n)}, b_{\pm}^{(n)}$ and $\gamma_{k_f}$ is the phase of $\CG_R(k_f, \omega)$.
This Green's function has a pole in the complex $\omega$ plane at 
\be
\omega_c  =  \omega_*(k) -i \Gamma(k) 
= \bca \left({k_\perp \over h_2}\right)^{1 \over 2 \nu_{k_f}} e^{- i {\gamma_{k_f} \over 2 \nu_{k_f}}} & \nu_{k_f} < \half \cr
      v_f k_\perp - v_f h_2 e^{i \gamma_{k_f}} (v_f k_\perp)^{2 \nu_{k_f}} & \nu_{k_f} > \half
\eca  .
%
 \label{dispersion}
\ee
We interpret the $\omega=0$, $k_\perp=0$ singularity as a Fermi surface and the finite $\omega$ poles as particle-like excitations above this Fermi surface. 
Looking at (\ref{dispersion}), the excitations have dispersion relation $\omega_*(k) \propto k_\perp^z$ 
and widths $\Gamma(k) \propto k_\perp^\delta$ 
with
\be
z = \bca {1 \over 2 \nu_{k_f}} & \nu_{k_f} < \half \cr
            1 & \nu_{k_f} > \half
            \eca
 \label{dispersionone}
 ~~~~~\text{and}~~~~~
\delta = \bca {1 \over 2 \nu_{k_f}} & \nu_{k_f} < \half \cr
            2 \nu_{k_f} & \nu_{k_f} > \half
            \eca \ .
\ee
For $\nu_{k_F}<\frac{1}{2}$, the width and energy are comparable and the excitations are not stable quasi-particles; these are non-Fermi liquids.  For $\nu_{k_F}>\frac{1}{2}$, as we scale towards the Fermi surface, the ratio of lifetime to energy goes to zero and these are stable particles. For $\nu_{k_F}= 1/2$, both $\CG_R^\alpha (\omega)$ and $a_{+ \alpha}^{(1)}$ have poles which cancel, leaving a log in the Green's function. It is
\be
G_R^\alpha(\omega, k)\approx \frac{h_1}{k_\perp+ c_1  \omega+ \tilde{c}_1 \omega \log \omega}
\ee
with $\tilde{c}_1$ real and $c_1$ complex. 
Such a Green's function 
is characteristic of 
``marginal Fermi liquids," 
which provide a phenomenological model of the strange metal state of the cuprates \cite{varma}.

Thus, we have obtained Green's functions for a family of excitations about a Fermi surface. The low energy properties, such as the form of the lifetime and dispersion relation are entirely determined by the scaling dimensions of an emergent conformal field theory.

\section{Turning on Dipole Couplings}
We want to look at the effects of
changing the intrinsic electric or magnetic dipole moment of the bulk spinor
 on the structure of these holographic Fermi surfaces. To do this, we use the bulk Lagrangian density
\be
\CL= i (\bar{\Psi} \Gamma^M D_M \psi - m \bar{\psi} \Psi) - \bar{\Psi} (g_m + g_e \Gamma) \Gamma^{MN} \Psi  F_{MN}
\ee
with (in our basis \eqref{eq:Diracbasis})
\be
\Gamma \equiv \Gamma^{\underline{t}} \Gamma^{\underline{r}} \Gamma^{\underline{x}} \Gamma^{\underline{y}}=\open 0 &- i\sigma^2  \\  -i \sigma^2 & 0 \close
\ee
The Dirac equation is now
\be
\left( \Gamma^{M} D_M  - m  + i (g_m + g_e \Gamma) \Gamma^{MN} F_{MN} \right) \Psi=0
 \ee
Once again, we can cancel the spin part of the covariant derivative by making the definition (\ref{fieldredefinition}). 
Using $F_{rt}= \mu r_0/r^2$, and the definition \eqref{fieldredefinition}, with 
$\psi=\left( \begin{array}{c} \Phi_1 \\ \Phi_2 \end{array} \right)$
in the basis \eqref{eq:Diracbasis}, we get
\bea
(\partial_r + m \sqrt{g_{rr}} \sigma^3 ) \Phi_\alpha&=& \sqrt{ \frac{g_{rr}}{-g_{tt}}} \left( \omega + \mu_q (1-\frac{r_0}{r}) \right)  i \sigma^2 \Phi_\alpha + \sqrt{ \frac{g_{rr}}{g_{ii}}} k (-1)^\alpha \sigma^1 \Phi_\alpha \non \\ 
& + & 2 \mu r_0 (-g_{tt})^{-\frac{1}{2}} (g_m \sigma^1 \Phi_\alpha +  g_e \sigma^3 \Phi_\beta) \frac{1}{r^{2}} \label{dipoleDirac}
\eea
where again $\alpha=1,2$ and the index $\beta \equiv 3 - \alpha$ awkwardly indicates the other component. Note that 
when $g_e \neq 0$,
the Dirac equation is no longer block diagonal in this basis, though it is still real. The dipole terms have no effect on the boundary behavior of this equation.

However, because there is mixing when $g_e \neq 0$, the process for extracting the Green's function is slightly more complicated. Equation (\ref{matrixe}) still holds, but we can no longer choose two sets of boundary conditions such that $G_R$ is diagonal. Instead, we use two sets of linearly independent boundary conditions, $I$, and $II$. (\ref{matrixe}) becomes
\be
\open b_1^I & b_1^{II} \\ b_2^I & b_2^{II} \close = \open s_1 & s_2 \\ s_3 & s_4 \close \open a_1^I &  a_1^{II} \\ a_2^I & a_2^{II} \close  \label{abdef}
\ee
or $B=SA$ in matrix notation. The Green's function is $G_R(\omega, k)=S = B A^{-1}$. 

\subsection{Discrete Symmetries}
We can discover several discrete symmetries by examining the effect of conjugating the Dirac equation (\ref{dipoleDirac}) and the infalling boundary conditions with certain simple matrices, $U$. For example, when our two sets of infalling boundary conditions correspond to the two different spins, conjugating with the matrix 
\be
U= \open 0 & 1 \\ 1 & 0 \close=i\Gamma^{\underline{x}} \Gamma^{\underline{y}}
\ee
switches the sign of $k$ in the Dirac equation and switches the two sets of boundary conditions. We learn that
\be
G(\omega, -k) = U G(\omega, k) U
\ee
When $g_e=0$, we can take a diagonal basis,  leading to $G_1(\omega, -k)= G_2(\omega, k)$. For the general mixed case, we note that $\det G(\omega, -k) = \det G(\omega, k)$, so that our graphs of Fermi surfaces in the $(k,q)$ plane will be invariant under $k \rightarrow -k$.  In a similar way, by examining the effect of $U$ on (\ref{dipoleDirac}) and on the boundary conditions,  the choice
\bea
U= \open \sigma^3 & 0 \\ 0 & \sigma^3 \close = \Gamma^{\underline{r}}& \Longrightarrow & G(-\omega, -k ; -q, -g_m, g_e)=-G^*(\omega, k;q, g_m, g_e) ~.
\eea
This, along with the first discrete symmetry,  implies that our Fermi surface plots with $g_m=0$ will be symmetric under $q \rightarrow -q$. Finally, the choice
\bea
U= \open \sigma^2 & 0 \\ 0 & \sigma^2 \close=\Gamma^{\underline{r}}\Gamma^{\underline{t}}& \Longrightarrow & G(\omega, -k ; -m, -g_m, -g_e)=-\left[G(\omega, k; m , g_m, g_e)\right]^{-1} ~.
\non \\
\eea
In particular, this implies that switching to alternative quantization is equivalent to taking $(m, g_m, g_e) \rightarrow (-m, -g_m, -g_e)$. 

\subsection{The Low Frequency Limit} 
Again we develop a perturbation series in $\omega$ by splitting the $r$-axis into inner and outer regions. The lowest order inner region equation is
\bea
(-\partial_\zeta + \frac{m R_2}{\zeta}  \sigma^3 ) \Phi_\alpha&=& \left( \omega +\frac{q e_3}{\zeta} \right) i \sigma^2 \Phi_\alpha +  \frac{R_2}{r_0 \zeta} k (-1)^\alpha \sigma^1 \Phi_\alpha \non \\ 
& + & 2 \frac{e_3}{\zeta}  (g_m \sigma^1 \Phi_\alpha+  g_e \sigma^3 \Phi_\beta) \frac{1}{R_2}~. \label{adstwodip}
\eea
Near the boundary of $AdS_2$, we get
\be
-\zeta \partial_\zeta \Phi_\alpha=i \sigma_2 q e_3\Phi_\alpha-R_2\left(m \sigma_3+ \tilde{M}_\alpha\sigma^1 \right) \Phi_\alpha+2 g_e \frac{e_3}{R_2}\sigma^3 \Phi_\beta
\ee
with
\be
\tilde{M}_\alpha \equiv -(-1)^\alpha \frac{ k}{r_0}  +2 \frac{e_3}{R^2_2} g_m ~~~.
\ee
Again
$
-\zeta \partial_\zeta \psi=U(g_e, g_m) \psi
$
and the four eigenvalues of $U$ are
\be
\pm \nu_{1,2}=\pm \frac{1}{R_2} \sqrt{(m^2+\frac{k^2}{r_0^2})R_2^4+ e_3^2  (4(g_e^2+g_m^2)-q^2R_2^2) \pm 4 e_3 R_2^2 \sqrt{g_m^2 \frac{k^2}{r_0^2} + g_e^2 (m^2+\frac{k^2}{r_0^2})}}   \label{eigenvalues}
\ee
where the $1,2$ correlates with the $\pm$ in the square root. Thus the dimensions of operators in the IR CFT are significantly changed. In the case of $g_e=g_m=0$, the usual case of two degenerate eigenvalues obtains. 

By making a basis change on (\ref{adstwodip}), we can block diagonalize it (though we cannot do so for the full Dirac equation):
\be
\left(-\partial_\zeta - i (\omega + \frac{q e_3}{\zeta}) \sigma_2+ \frac{\nu_{\pm}}{\zeta} \sigma_1 \right) \tilde{\Phi}_{1,2}=0
\ee
with
\bea
\nu_- & = & \sqrt{\frac{4 e_3^2 \left(g_e^2+g_m^2\right) R_2^2+\left(m^2+\frac{k^2}{r_0^2}\right)
   R_2^6-4 \sqrt{e_3^2 \left(g_m^2 \frac{k^2}{r_0^2}+g_e^2
   \left(m^2+\frac{k^2}{r_0^2}\right)\right) R_2^8}}{R_2^4}} \non \\
\nu_+ & = & \sqrt{\frac{4 e_3^2 \left(g_e^2+g_m^2\right) R_2^2+\left(m^2+\frac{k^2}{r_0^2}\right)
   R_2^6+4 \sqrt{e_3^2 \left(g_m^2 \frac{k^2}{r_0^2}+g_e^2
   \left(m^2+\frac{k^2}{r_0^2}\right)\right) R_2^8}}{R_2^4}}  \non \\
\eea
This is exactly the same $AdS_2$ Dirac equation as (\ref{adsDirac}), after the replacement
\bea
(-1)^\alpha \frac{k}{r_0} & \rightarrow & -\frac{\nu_{\pm}}{R_2} \non \\
m & \rightarrow & 0 \non \\
\nu_\alpha & \rightarrow & \sqrt{\nu_{\pm}^2-q^2 e_3^2}=\nu_{1,2}~.
\eea
Thus,
\be 
\CG_R^{1,2}(\omega) = e^{-i \pi \nu_{1,2}} \frac{\Gamma(-2 \nu_{1,2}) \Gamma(1 + \nu_{1,2} - i q e_3)}{\Gamma(2 \nu_{1,2}) \Gamma( 1 - \nu_{1,2} - i q e_3) } \times \frac{-i \nu_{\pm} - i q e_3 - \nu_{1,2}}{-i \nu_{\pm} - i q e_3 + \nu_{1,2}} ( 2 \omega) ^{2 \nu_{1,2}} ~.\label{Green'sbaby}
\ee
As in (\ref{thematching}), we can match in the outer region onto either $\CG^1_R(\omega)$ or $\CG^2_R(\omega)$, this defines our two boundary conditions. The components in the outer region, however, will generically be mixed $4$-spinors. We will have two solutions
\be
\psi^I  = \eta_+^{I} + \CG^1_R (\omega)  \eta_{-}^{I}  ~~~~
\psi^{II} = \eta_+^{II} +  \CG^2_R (\omega)  \eta_{-}^{II}  
\ee
We can expand the matrices $A$ and $B$ (\ref{abdef})  perturbatively in $\omega$ near the boundary. For example, 
\be
B= B_+^{(0)} + \omega B_+^{(1)} + O(\omega^2) + (B_-^{(0)} + \omega B_-^{(1)} + O(\omega^2)) \CG_R(\omega)
\ee
with 
\be
\CG_R(\omega) \equiv \open \CG^1_R (\omega) & 0 \\ 0 & \CG^2_R(\omega) \close
\ee
The equation for the low frequency Green's function is (to order $\omega^2$)
\be
B_+^{(0)} + \omega B_+^{(1)}  + (B_-^{(0)} + \omega B_-^{(1)}  ) \CG_R(\omega)=G_R (\omega, k) \left [( A_+^{(0)} + \omega A_+^{(1)} ) + (A_-^{(0)} + \omega A_-^{(1)} ) \CG_R(\omega) \right] ~.\\
\ee
All previous equations for correlation functions (such as (\ref{fscorr})) hold, with $a_\alpha$'s and $b_\alpha$'s replaced by matrices $A$ and $B$, $\CG_R(\omega)$ replaced by the matrix \eqref{Green'sbaby}, and all denominators replaced by matrix inverses. The Fermi surface is now defined by 
\be
\det [A_+^{(0)}(k_f)]=0~.
\ee 
The dispersion relation and width -- the analogues of (\ref{dispersionone}) --  are determined by solving 
\be
\det \left [(A_+^{(0)} (k_f) + \partial_k A_+^{(0)} (k_f) k_\perp  + \omega A_+^{(1)}(k_f ) + A_-^{(0)}(k_f) \CG_R(\omega) \right] =0~.
\ee
Thus, in general, the dispersion relation and width will be controlled by the smallest of $\nu_1$, $\nu_2$. 

For simplicity, we will find it easiest to deal with nonzero $g_m$ and $g_e$ separately. 

\subsection{$g_m \neq 0$, $g_e=0$}

For $g_e=0$, there is no need to do any basis changing; the Dirac equation is block diagonal. The near horizon equation is (setting $r_0=1$)
\be
(-\partial_\zeta + \frac{m R_2}{\zeta} \sigma_3) \Phi_\alpha=(\omega+  \frac{q e_3}{\zeta}) i \sigma_2 \Phi_\alpha+ \frac{R_2}{\zeta} ( (-1)^\alpha k + \frac{2 e_3 g_m}{R_2^2}) \sigma_1 \Phi_\alpha ~.\label{basischange}
\ee
Thus the effect of the magnetic dipole in the near horizon limit is to shift the momentum oppositely in the two blocks.  In the matching region (the $AdS_2$ boundary), the solution goes like 
\be
\Phi_\alpha \sim \zeta^{-\nu_\alpha} v_{+ \alpha} + \zeta^{\nu_\alpha} v_{- \alpha}
\ee
with
\be
\nu_\alpha= \frac{1}{R_2} \sqrt{(k R_2^2+ (-1)^\alpha 2 g_m e_3)^2+m^2 R_2^4 - q^2 e_3^2 R_2^2} \label{magneticdimension}
\ee
and
\be
v_{\pm \alpha} = \left( \begin{array}{c} m R_2 \mp \nu_\alpha \\ e_3 (q - \frac{2g_m}{R_2}) - (-1)^\alpha k R_2 \end{array} \right)~.
\ee
Matching onto the near horizon region, 
\be
\Phi_\alpha^{O}=\eta_{+ \alpha}+  \CG_R^\alpha(\omega) \eta_{- \alpha}
\ee
where the effect of the dipole coupling is to shift $k$ in $\CG_R^\alpha(\omega)$, and to change the UV data $\eta_{\pm}$. The $AdS_2$ Green's function is
\bea
\CG_R^{\alpha}(\omega) & = & e^{-i \pi \nu_\alpha} \frac{\Gamma(-2 \nu_\alpha) \Gamma(1 + \nu_\alpha- i q e_3)}{\Gamma(2 \nu_\alpha) \Gamma( 1 - \nu_\alpha- i q e_3) } \times \frac{(m + i n_\alpha) - i q e_3 -\nu_\alpha}{(m + i n_\alpha)- i q e_3 +\nu_\alpha} ( 2 \omega) ^{2 \nu_\alpha} \non \\
n_\alpha & = & (-1)^\alpha k + \frac{2 e_3 g_m}{R_2^2}
\eea

\subsection{$g_e \neq 0$, $g_m=0$}
In this case, the ($r_0=1$) near horizon equation can be block-diagonalized to the form\footnote
{
In the basis \eqref{eq:Diracbasis}, the Dirac equation takes the form
\be
(-\partial_\zeta + \frac{m R_2}{\zeta} \sigma_3) \Phi_\alpha=(\omega+  \frac{q e_3}{\zeta}) i \sigma_2 \Phi_\alpha+ \frac{R_2}{\zeta}  (-1)^\alpha k \sigma_1 \Phi_\alpha + \frac{2 e_3}{\zeta} g_e \sigma^3 \Phi_\beta 
\ee
The unitary transformation which block diagonalizes this is $\Phi = U \tilde \Phi$
\bea
U &= & \frac{1}{2 \sqrt{2} \sqrt{ k^2 + m^2 + m \sqrt{m^2 + k^2}}} \left( \begin{array}{cccc} -A_+&  -A_- &i A_- & -i A_+ \\ A_- & -A_+ & i A_+ & i A_- \\ -A_- & -A_+ & -i A_+ & i A_- \\ A_+ & -A_- & -i A_- & -i A_+ \end{array} \right) \non \\
A_\pm & = & m \pm k + \sqrt{m^2 + k^2}~~.
\eea
}
\be
-\partial_\zeta  \tilde{\Phi}_\alpha=(\omega+  \frac{q e_3}{\zeta}) i \sigma_2 \tilde{\Phi}_\alpha+ \frac{R_2}{\zeta} (  (-1)^\alpha \sqrt{m^2 + k^2}+ \frac{2 e_3 g_e}{R_2^2}) \sigma_1 \tilde{\Phi}_\alpha~~. \label{basischange2}
\ee
In the matching region, the solution again goes like 
$
\tilde{\Phi}_\alpha \sim \zeta^{-\nu_\alpha} v_{+ \alpha} + \zeta^{\nu_\alpha} v_{- \alpha}
$
with
\bea
\nu_\alpha &=& \frac{1}{R_2} \sqrt{(\sqrt{m^2 + k^2} R_2^2+ (-1)^\alpha 2 g_e e_3)^2 - q^2 e_3^2 R_2^2}  \\  \non
v_{\pm \alpha} &=& \left( \begin{array}{c} \mp \nu_\alpha \\ e_3 (q - \frac{2g_e}{R_2}) - (-1)^\alpha \sqrt{m^2 + k^2} R_2 \end{array} \right)~~.
\eea
The full outer region solution is $
\tilde{\Phi}_\alpha^{O}=\eta_{+ \alpha}+  \CG_R^\alpha(\omega) \eta_{- \alpha}
$
with
\bea
\CG_R^{\alpha}(\omega) & = & e^{-i \pi \nu_\alpha} \frac{\Gamma(-2 \nu_\alpha) \Gamma(1 + \nu_\alpha- i q e_3)}{\Gamma(2 \nu_\alpha) \Gamma( 1 - \nu_\alpha- i q e_3) } \times \frac{i n_\alpha - i q e_3 -\nu_\alpha}{ i n_\alpha- i q e_3 +\nu_\alpha} ( 2 \omega) ^{2 \nu_\alpha} \non \\
n_\alpha & = & (-1)^\alpha \sqrt{m^2 + k^2} + \frac{2 e_3 g_e}{R_2^2}~.
\eea

\section{Numerical Results}
To find Fermi surfaces, we look for $k_f$ such that $a_+^{(0)}(k_f)=0$. By (\ref{allbdy}), this corresponds to $\omega=0$ solutions to the Dirac equation which are normalizable (because of mixing, the process is slightly more complicated for $g_e\neq 0$; we review it below). We implement this procedure by numerically integrating the $\omega=0$ equation to the boundary and looking for zeros of $a_+^{(0)}$ for some range of $k$ and $q$. 

\begin{figure}
\centering
\includegraphics[scale=.5]{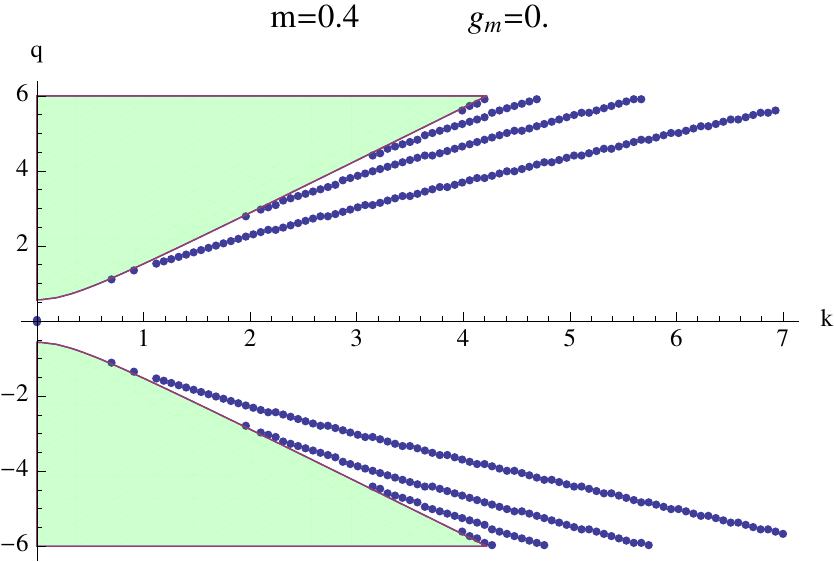} 
\caption{Fermi Surfaces $g_e=g_m=0, m=0.4$}
\label{nocouplings}
\end{figure}

Such numerical work was previously done\cite{Liu:2009dm, Faulkner:2009wj} for $g_m=g_e=0$ . There, it was found that Fermi surfaces existed in branches in the $(k,q)$ plane that were basically straight lines jutting out of an oscillatory region (a region where the $AdS_2$ operator dimensions (\ref{firsteigen}) are imaginary and inside which there exist no Fermi surfaces). See Figure \ref{nocouplings} for such a graph with $m=0.4$. The oscillatory region is shaded green. 

\subsection{$g_m \neq 0$, $g_e =0$}
For $g_m \neq 0$, the above structure is preserved; there are Fermi surface branches jutting out of oscillatory regions in the $(k,q)$ plane. By (\ref{magneticdimension}), turning on $g_m$ keeps intact the shape of the oscillatory region, moving it to larger $k$ (it also moves another copy 
associated with the opposite spin
to smaller $k$, but we focus on $k>0$ as everything is $k\rightarrow -k$ invariant). We make the following qualitative observations

\begin{enumerate}
\item{As the oscillatory region moves to larger $k$, it ``eats" Fermi surfaces in the $(k,q)$ plane.  These Fermi surfaces branches move to higher $q$ for larger $|g_m|$ (see Figure \ref{gmprogress}).

\begin{figure}[h]
\centering

\includegraphics[scale=0.5]{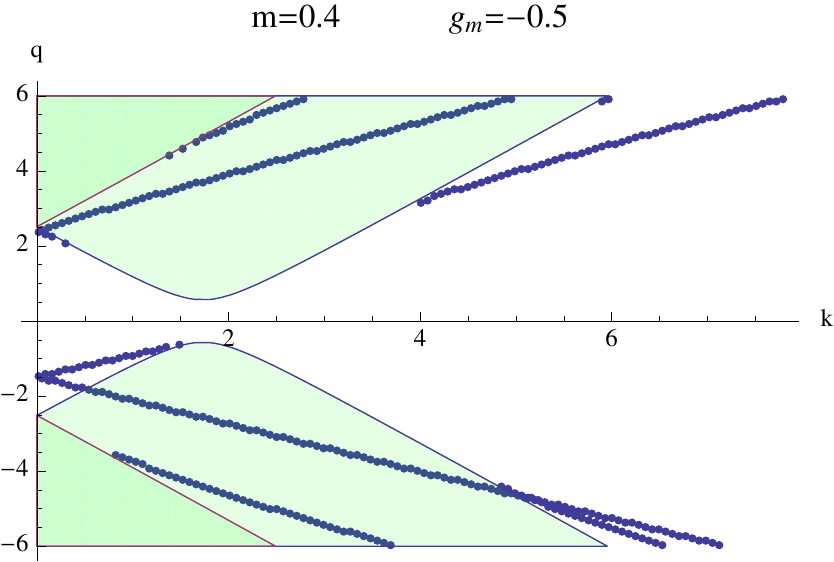}
\includegraphics[scale=0.5]{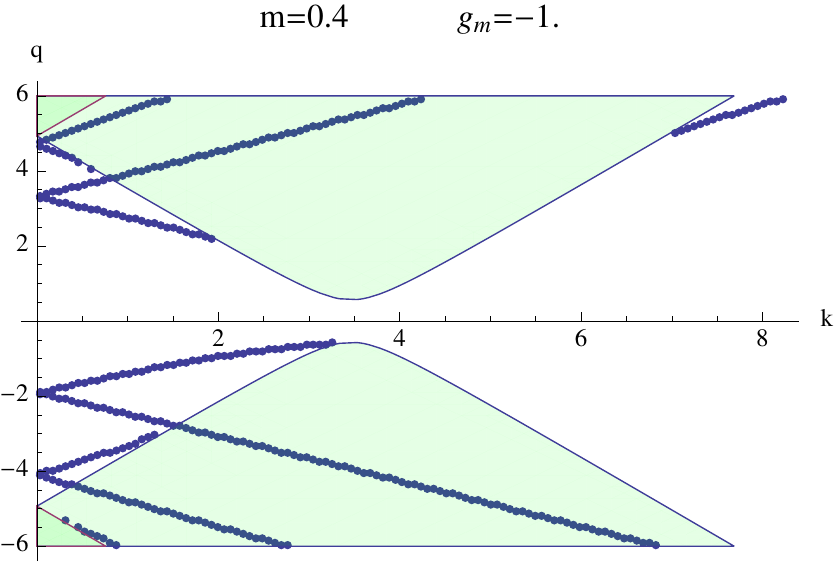}
\includegraphics[scale=0.5]{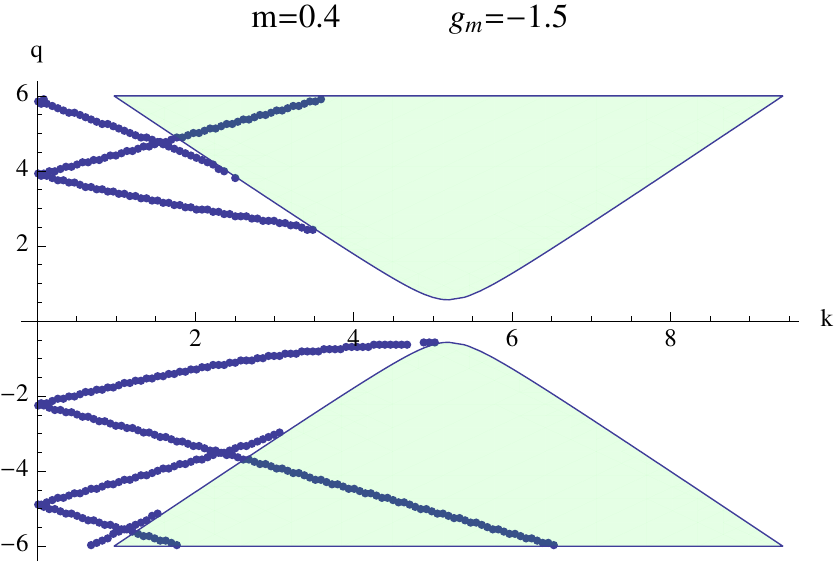} 
\includegraphics[scale=0.5]{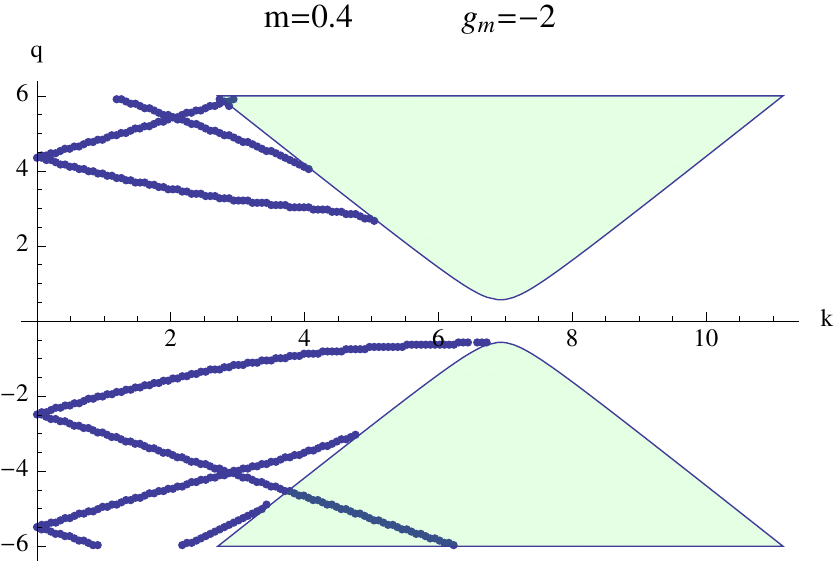} 
\caption{Fermi Surfaces for Increased $|g_m|$ with $m=0.4$}
\label{gmprogress}
\end{figure}

}

\item{The dipole coupling has the most effect at low $q$, where it flattens and curves Fermi surface branches close to the oscillatory region. Far from the oscillatory regions, the branches asymptote to straight lines. 
}

\item{This effect is most pronounced for $m<0$ (alternative quantization). For $m$ negative enough, local and global maxima and minima can develop in Fermi surface branches near the oscillatory region. See Figure \ref{min} for an example of how such a minimum develops as $m$ is lowered. Also, in Figure \ref{fixedgm} we plot results for $g_m$ fixed and $m=-0.4, 0, 0.4$.

}

\end{enumerate}

\begin{figure}[h]
\centering
\includegraphics[scale=.5]{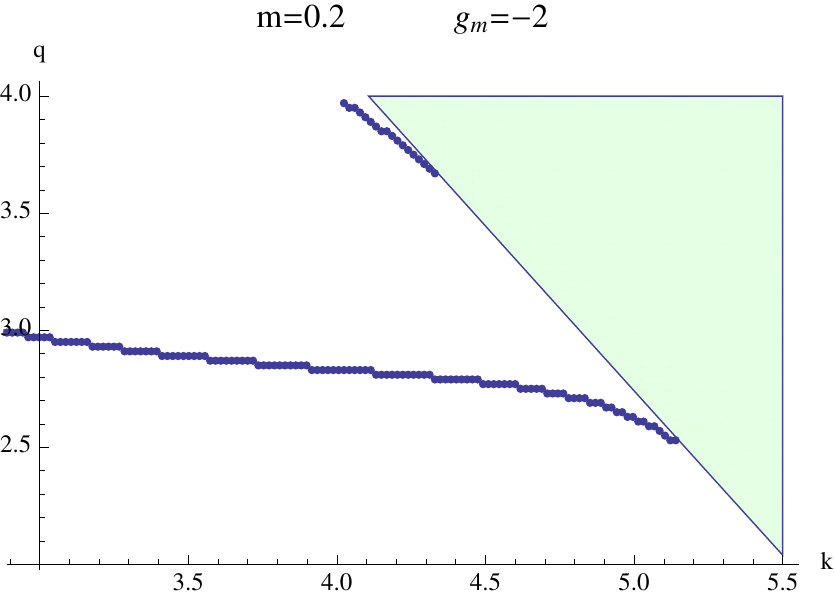}
\includegraphics[scale=.5]{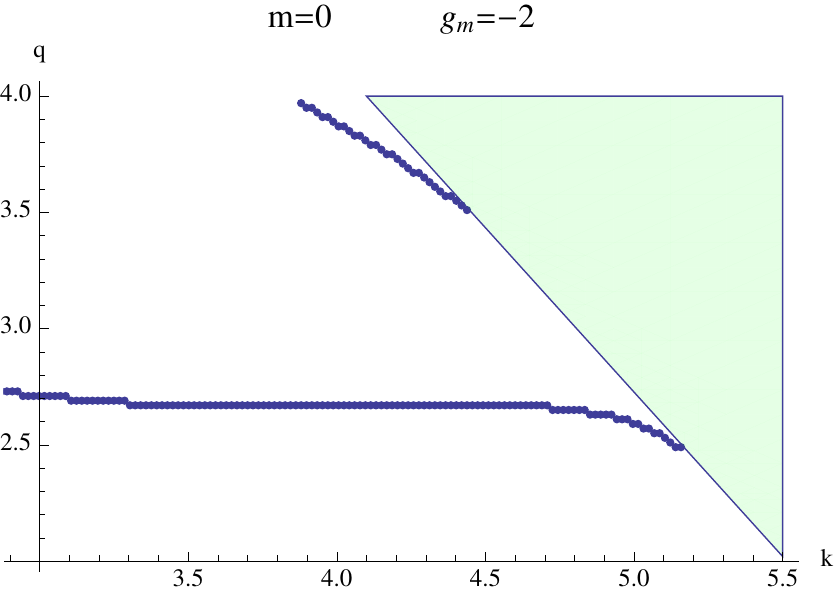} 
\includegraphics[scale=.5]{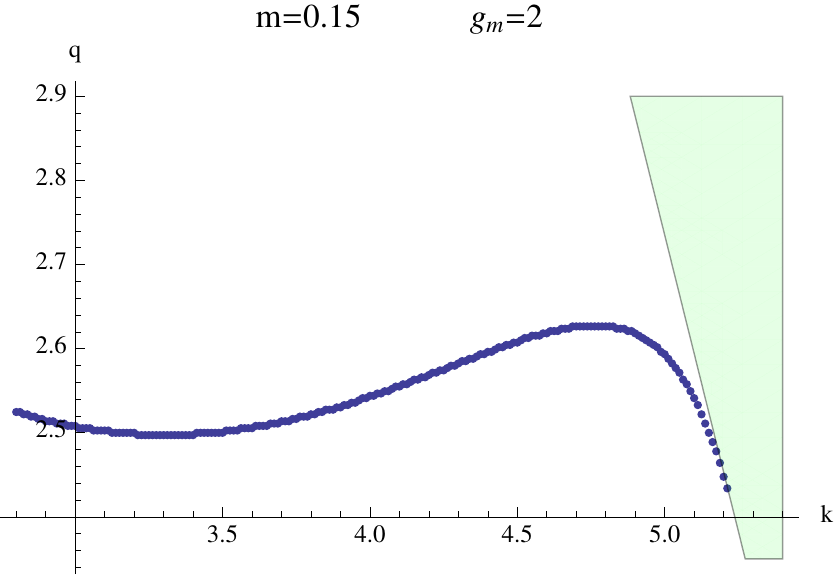} 
\includegraphics[scale=.5]{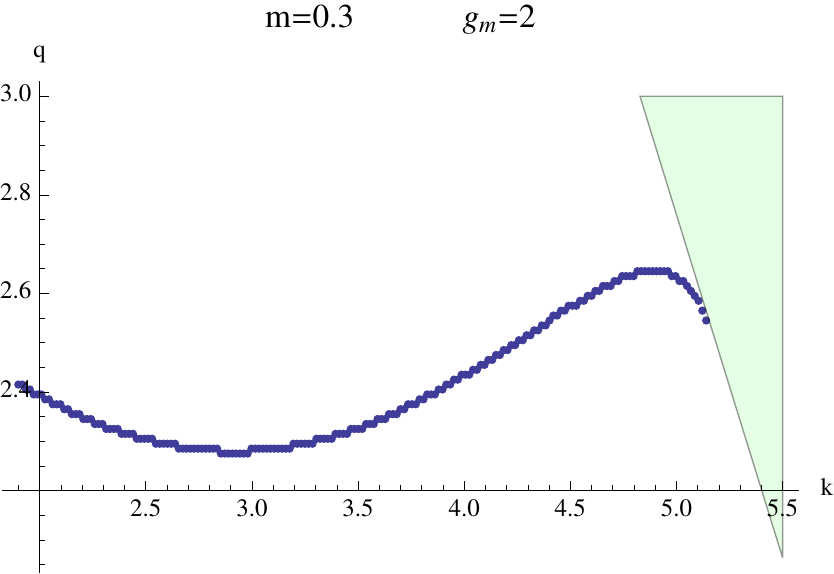} 
\caption{Developing local and global minimum for $m<0$ with $g_m=-2$}
\label{min}
\end{figure}

\begin{figure}[h]
\centering
\includegraphics[scale=.4]{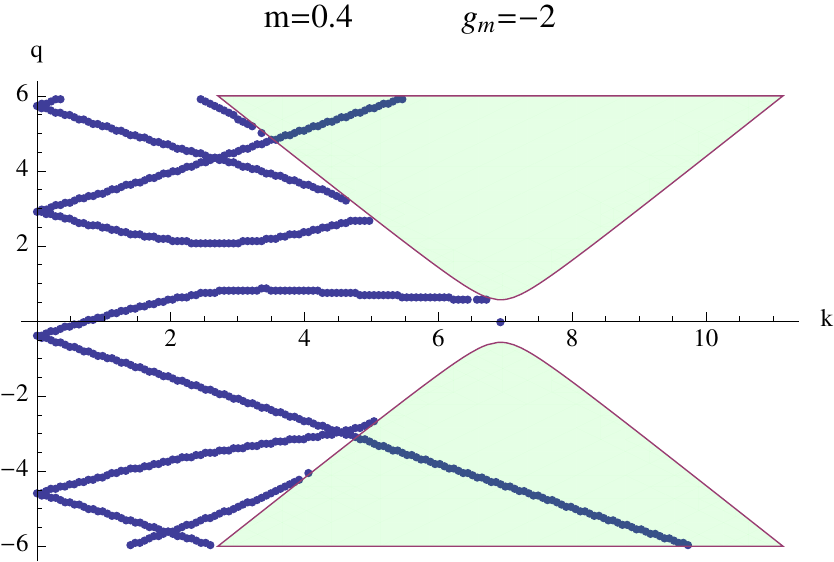}
\includegraphics[scale=.4]{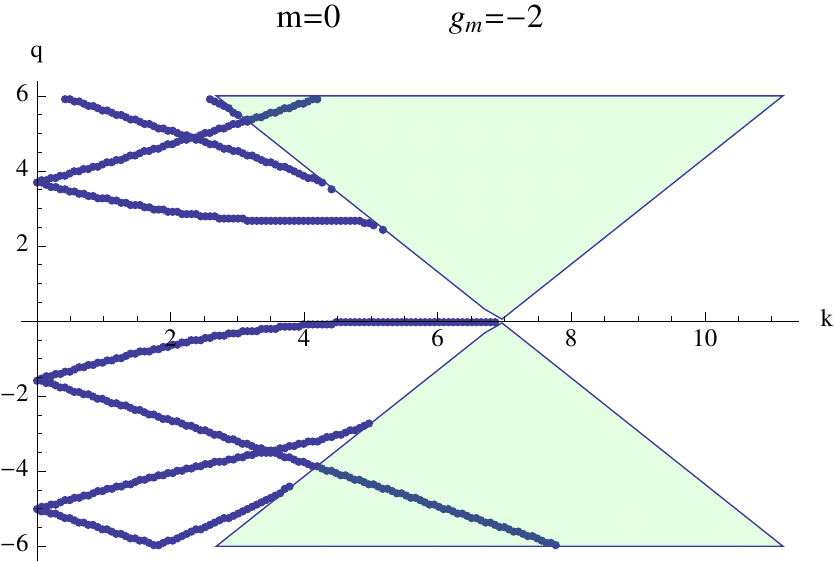} 
\includegraphics[scale=.4]{gmtwo.pdf} 
\caption{Fixed $g_m=-2$ for $m=-0.4, 0, 0.4$}
\label{fixedgm}
\end{figure}

In \citelow{Faulkner:2009wj}, a ``phase diagram" was constructed in the $(m,q)$ plane which showed the attainable $\nu_\alpha$'s for the primary Fermi surface (that with the largest $k_f$ for a given $q$). Here, we construct a similar phase diagram for $g_m=-2$. Because of various ambiguities
that arise when there are multiple Fermi surfaces, we focus on a single $q>0$ branch. 
Because the branch gets flattened near the oscillatory region, there are more attainable $\nu$'s than with $g_m=0$ for the same range of $m,q$. Also, because of point 3 above, when global or local minima occur we must pick what we mean as the primary Fermi surface within a branch (note that this differs from $g_m=0$ where the choice is made between \emph{different} branches). We choose that Fermi surface with the largest $\nu_\alpha$. Note, in this case, such a Fermi surface actually has smaller $k_f$.

\begin{figure}[h]
\centering
\includegraphics[scale=0.3]{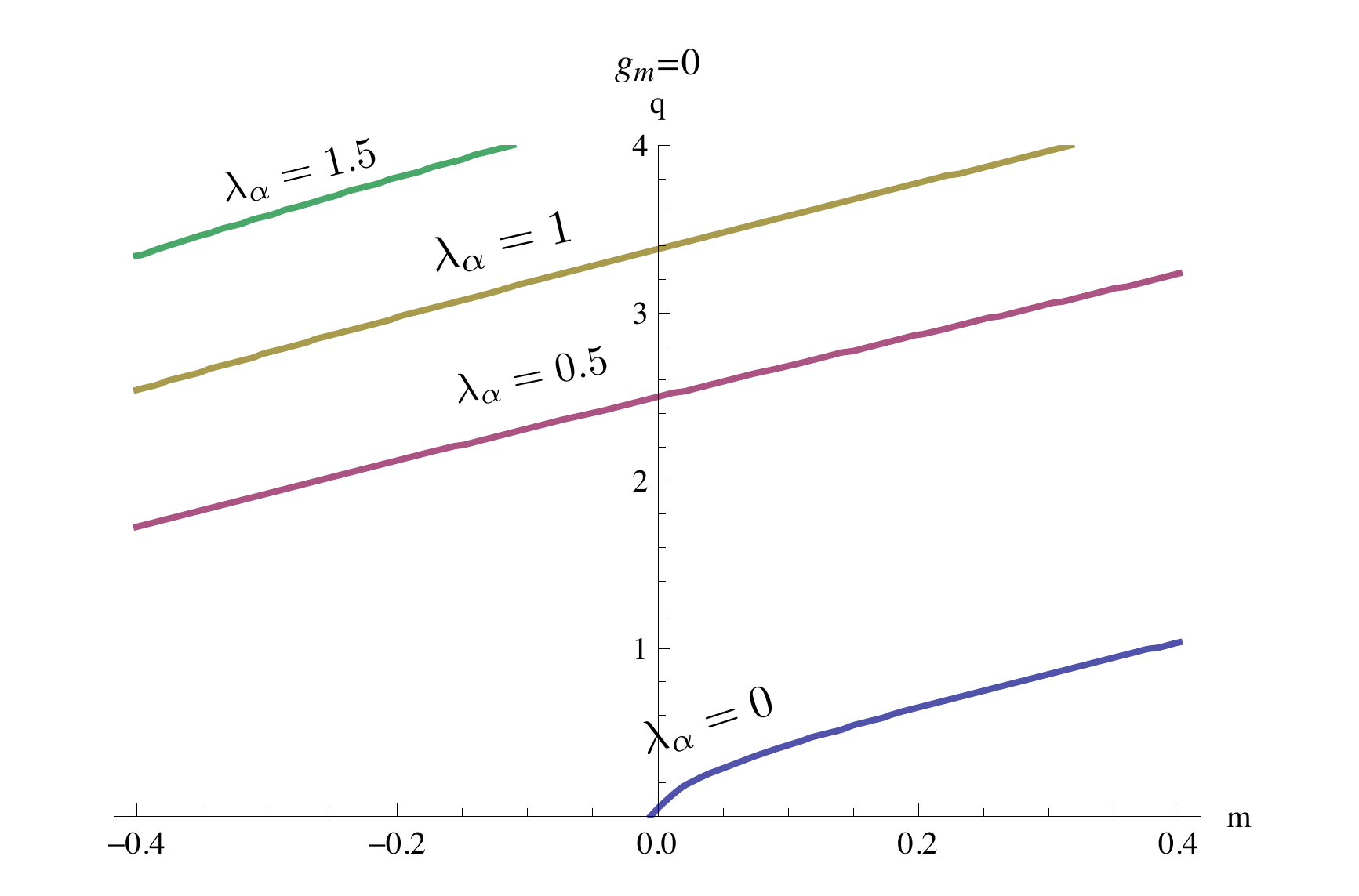}
\includegraphics[scale=0.3]{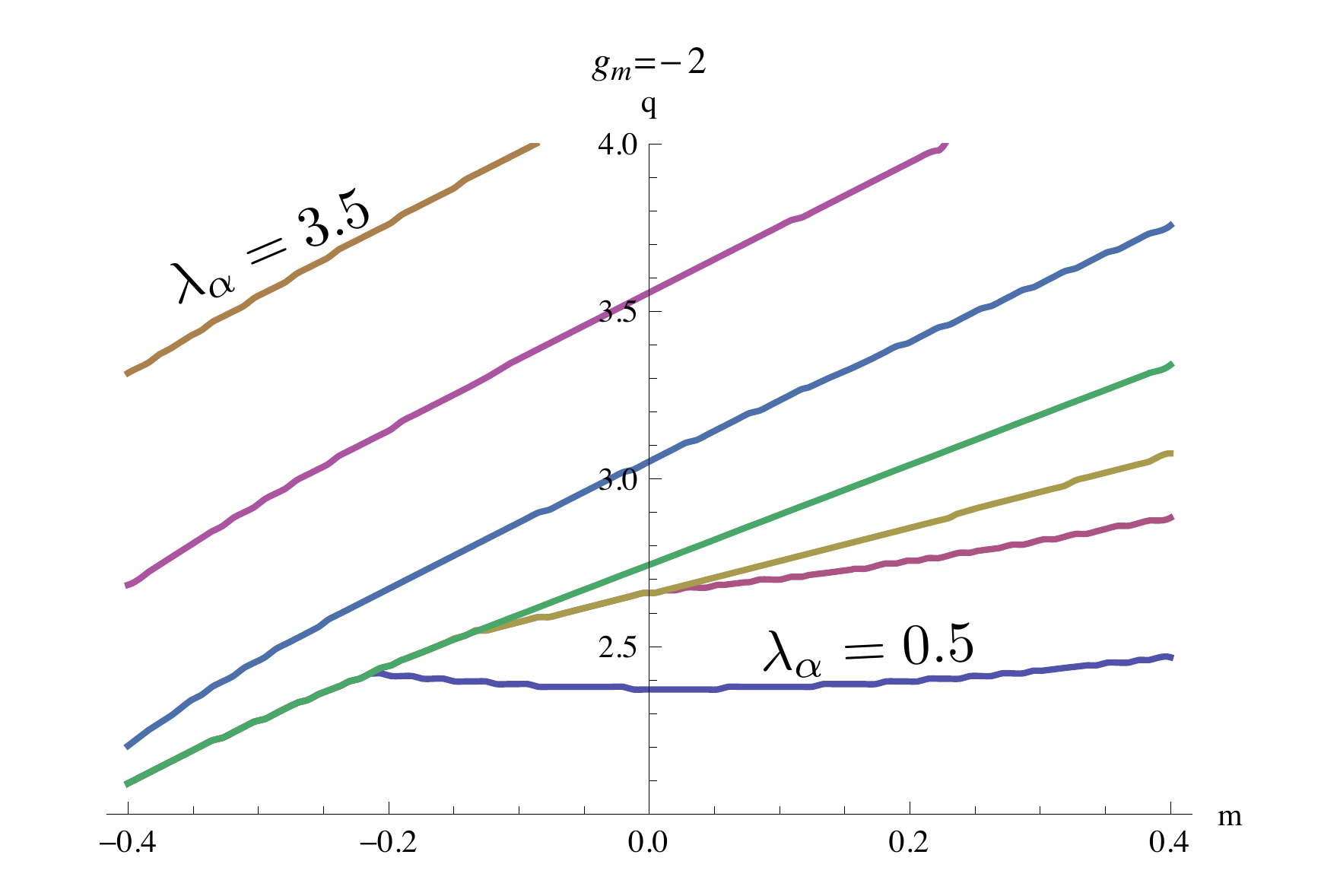}
\caption{``Phase Diagram" for $g_m=-2$ }
\end{figure}

\subsection{$g_e \neq 0$, $g_m=0$}
For $g_e \neq 0$, there is mixing between the spin components. We change bases in the $\omega=0$ Dirac equation so that  (\ref{basischange2}) is the near horizon limit. We then use two different infalling boundary conditions, each corresponding to a distinct $AdS_2$ dimension. We integrate this out to the boundary, change basis back to the original spin basis and numerically look for zeros of $\det A_+^{(0)}=a_{+1}^{I (0)} a_{+2}^{II (0)}-a_{+1}^{II(0)} a_{+2}^{I(0)}$ (see (\ref{abdef})). Some observations:

\begin{figure}[h]
\centering
\includegraphics[scale=.5]{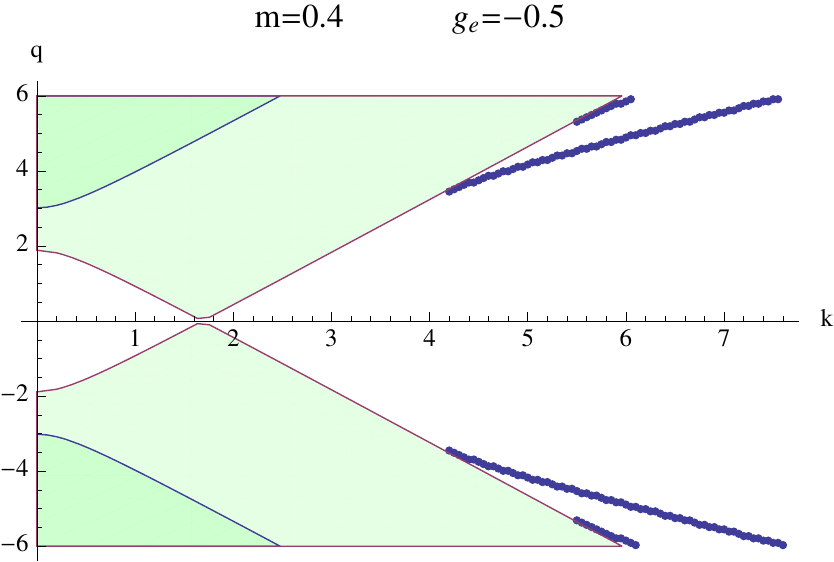}
\includegraphics[scale=.5]{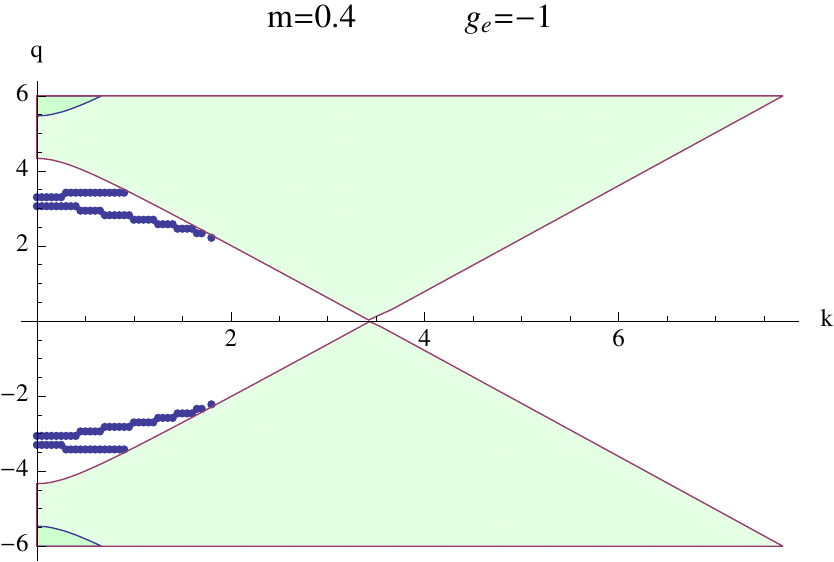}
\includegraphics[scale=.5]{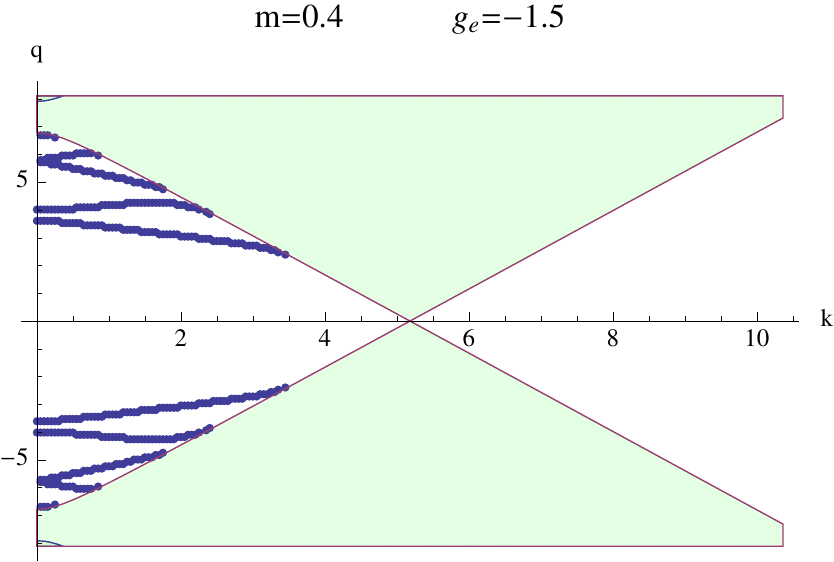} 
\includegraphics[scale=.5]{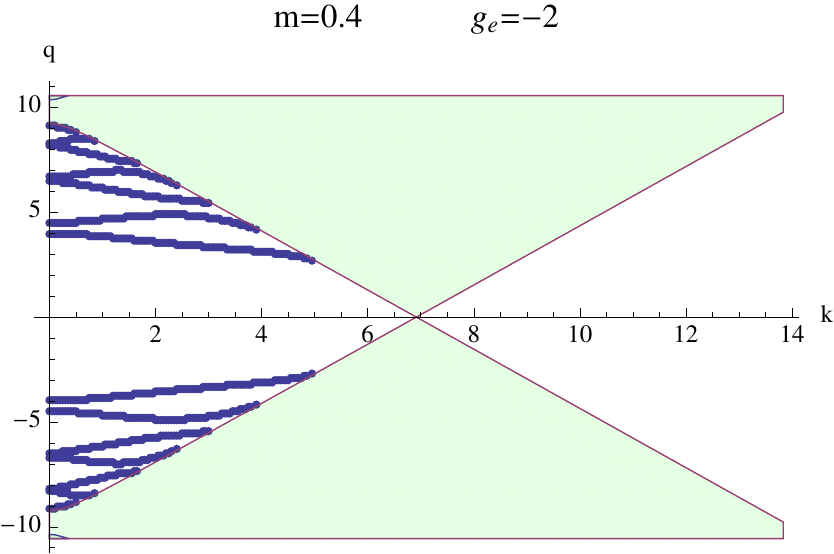} 
\caption{Increasing $|g_e|$}
\label{geprogression}
\end{figure} 
 
\begin{enumerate}
\item{Fermi surface branches continue to jut out of an oscillatory region.  As $|g_e|$ is increased, the oscillatory regions kiss and move to the right (see figure \ref{geprogression}). As in the magnetic case, the oscillatory region ``eats" Fermi surface branches as it moves to the right.

}

\item{Fermi surface branches are created to the left (smaller $|k|$) of the oscillatory region, which we will call the ``interesting region". As one increases $|g_e|$ more Fermi surfaces are created in this region. There can also be local maxima or minima created near the oscillatory region as in the magnetic dipole case. 

}

\item{For $g_e \neq 0$ the Fermi surfaces are much more gently sloping in the interesting region than for Fermi surface branches with similar $g_m$ (and all other constants comparable). For $m=0$, the Fermi surface branches are nearly flat. 

}

\item{There seem to be small gaps between the Fermi surface branches at $k=0$. This indicates local maxima or minima at $k=0$. As one lowers $|m|$, the gaps become larger and larger, although for $m<0$ there always exists a branch with a large gap.

}

\item{Since $g_e$ mixes the spin components, it is difficult to represent 
the landscape of possible values of $\nu$ in the same way as previously.
}

\end{enumerate}





\section{Discussion}

We have found that 
the holographic framework 
for Fermi surfaces is robust under the change of the magnetic and electric dipole moments of the bulk spinor field, 
in the sense that the low-frequency Green's function remains
of the form determined in \citelow{Faulkner:2009wj}. 
Turning on these couplings moves the Fermi surfaces around,
and changes the scaling dimension of the emergent IR $AdS_2$ symmetry.

What is the meaning of the bulk dipole couplings in terms of 
properties of the fermion operator in the boundary field theory? In vacuum, these couplings do not affect the fermion two point function. They do, however, change the structure of the current-fermion-fermion three point function. The full calculation is complicated, but one simple characterization is the following. With $g_m=g_e=0$, there are no terms in $\vev{ \bar \Psi_\alpha \Psi_\beta J^\mu} $ proportional to second rank clifford algebra elements $\Gamma^{\mu\nu}$; turning on the dipole couplings creates such terms.  It would be interesting to understand better the physical significance of this.

It was previously found that the existence of Fermi surfaces in ordinary quantization was correlated with the existence of the oscillatory region in $k$ space where the $AdS_2$ scaling dimension becomes imaginary. The oscillatory region occurs for values of the momentum such that there is Schwinger pair production in the $AdS_2$ region.  A heuristic interpretation is that this creates a bulk Fermi surface, leading to a boundary one (the duality implies an equality of Hilbert spaces,
meaning that the bulk spectral density is related to the boundary spectral density; see \cite{Faulkner:2010da, resistivitypaper} for the precise relation between these two quantities). 
It appears that this mechanism involving pair production is not necessary to have a Fermi surface; boundary Fermi surfaces can appear -- in alternative quantization -- without oscillatory regions. 
However, the alternative quantization is unstable in the RG sense; any small addition of the double trace operator $\CO^\dagger \CO$ flows the CFT to that of ordinary quantization. Thus, the known Fermi surfaces without oscillatory region are also unstable; they flow away, as can be seen in Figure 5 of \citelow{Faulkner:2009wj}. 

The existence of an oscillatory region implies that the bulk Fermi sea has support at the black hole horizon. Naively, its gravitational backreaction should be suppressed by a factor of 
the Newton's constant, which is proportional to $1/N^2$. However, when one integrates the near-horizon charge density to some radial position $r$, there is a logarithmic divergence 
in $r$ which can offset this suppression beyond some critical $r_c \sim e^{ - N^2}$; 
for $r < r_c$ backreaction cannot be ignored \cite{Hartnoll:2009ns}. 
This backreaction was argued to change the geometry to a Lifshitz geometry \cite{Kachru:2008yh} 
with dynamical exponent $ z \sim N^2$.  This modifies the fermion response 
for frequencies and temperatures below some new low-energy scale $E_c \sim \mu e^{ - N^2}$
\cite{Faulkner:2010tq}.
Below these frequencies and temperatures, the behavior is that of a Fermi liquid (the self-energy is
analytic in frequency at the Fermi surface),
and therefore the system does not in fact describe a non-Fermi liquid groundstate.
We note that it is likely that other instabilities, such as the holographic superconductor
instability \cite{Gubser:2008px, Hartnoll:2008vx, Hartnoll:2008kx, Herzog:2009xv} set in at much higher temperatures.
The effects of such superconducting order on holographic Fermi surfaces
has been studied first for $s$-wave order \citelow{Chen:2009pt, Faulkner:2009am, Gubser:2009dt}
and more recently for order parameters with nodes \citelow{Gubser:2010dm, 
Ammon:2010pg, Benini:2010qc, Vegh:2010fc, Benini:2010pr}.

For the reason described in the previous paragraph, 
it would be interesting to find a deformation of the action which allows Fermi surfaces that are RG stable but do not come with an oscillatory region. Unfortunately, our results indicate that the dipole couplings are not such a deformation. Although the couplings change the shape and location of the oscillatory region in the $(k,q)$ plane, we again find that Fermi surfaces in ordinary quantization only occur with an oscillatory region. 

A subject of current interest is the form of the gravitational backreaction of the
density of spinor particles in the bulk \cite{Hartnoll:2009ns, Hartnoll:2010gu, Hartnoll:2010xj, Hartnoll:2010ik, 
Hung:2010te, vCubrovic:2010bf}.  
Given that the dipole couplings studied here change the shape of the Fermi surface and of the oscillatory region in momentum space, 
they will also have an interesting effect on the character of the back-reacted solutions\footnote
{We thank David Vegh for discussions of this issue.}.

Finally, we have found that the dipole operators curve Fermi surface branches in the $(k,q)$ plane close to the oscillatory region. For certain values of $g_m$ and $g_e$ we can create local maxima and minima of these branches. It would be interesting if we could embed this system into one where $q$ is a tunable parameter. In this context, a local maximum, for example, would represent two Fermi surfaces that merge and annihilate as $q$ is continuously increased. We leave such an embedding to future work.


{\bf Note Added 3:} 
The papers \citelow{Edalati:2010ww, Edalati:2010ge} attempt to interpret the 
motion 
(upon increasing $g_m$)
of the Fermi surface pole 
into the oscillatory region
and its subsequent evolution
as the formation of a gap\footnote
{JM would like to thank the authors of \citelow{Edalati:2010ww, Edalati:2010ge}
for correspondence which led 
to the following more careful consideration of these points.}.  
At first glance, this interpretation is problematic,
because a characteristic feature of the oscillatory region
is nonzero spectral weight at zero frequency 
(see figure 4 of \citelow{Liu:2009dm} and section IV.A of \citelow{Faulkner:2009wj}).  
At values of the dipole coupling when the Fermi surface first disappears, it 
enters the oscillatory region and the IR CFT exponent becomes $i$ times a small number;
the low-frequency spectral weight, while incoherent, is by no means small there.
The interpretation of
\citelow{Edalati:2010ww, Edalati:2010ge} 
of the formation of a gap 
relies on the numerical smallness of this weight 
at still-larger values of dipole coupling.

A partial explanation of this effect is the following. 
As the dipole coupling is increased further, for fixed $k$, 
one exits the oscillatory region again, and 
the IR CFT exponent \eqref{magneticdimension}
$\nu = \sqrt{ - q^2/2 + m^2L^2 + (k \pm c_d g_m)^2} $
becomes real and positive and, eventually, large.
Such a large, real IR CFT scaling dimension $\nu$
suppresses the incoherent spectral weight away from poles of the Green's function \citelow{Faulkner:2009am}.
This is because 
the low frequency spectral density satisfies \citelow{Faulkner:2009wj}
\be \Imm G_R \propto \left( { \omega \over \mu } \right)^{2 \nu } \ll 1 
\ee 
for $\omega < \mu$ and $\nu \gg 1$.
\footnote{
JM thanks the authors of  \citelow{Edalati:2010ww, Edalati:2010ge} for pointing out 
to him the similarity between figure 5 of 
\citelow{Faulkner:2009am}
and their results.}
\footnote{Note that the behavior seen in 
\citelow{Edalati:2010ww, Edalati:2010ge} and described here
should be distinguished from other mechanisms for suppression of low-frequency spectral weight 
observed in \citelow{Faulkner:2009am}. In the latter reference, 
two Fermi surface modes (in opposite-spin components) 
are pairing up as a result of the coupling to the superconducting order parameter.
Further, spectral weight at zero frequency is {\it exactly zero} in that case --
except for the FS pole, the weight vanishes everywhere outside the IR CFT lightcone.
In the situation studied in \citelow{Edalati:2010ww, Edalati:2010ge} and here, there is no pairing and no IR CFT lightcone
since $z = \infty$.
}
\footnote{
Yet another possible mechanism for the suppression of spectral weight can be envisioned from available information
about the low-frequency behavior of $G_R$ \cite{Faulkner:2009wj}, as follows;
we do not know how to implement this mechanism in a gravity model.
At low frequencies, the fermion spectral density
in the oscillatory region
may be written
\bea
\label{oscdentistry}
\rho_{{\rm osc}}(\omega, k ) = \Imm \frac{e^{ i \theta} |c| \omega^{ 2 i \lambda} + 1 }{e^{ i \theta'} |c| \omega^{2 i \lambda} + 1 }~~.
\eea
Recall that $ c \omega^{ 2i \lambda}$ is the IR CFT Green's function
(the IR CFT dimension is imaginary)
at $T=0$. $e^{ i \theta, \theta'}$ are scattering phases
constructed from the UV data.
This expression is valid in the oscillatory regime 
$ k < k_{{\rm osc}} $,
and is derived in \citelow{resistivitypaper}.
We observe that should the IR CFT dimension be $i$ times a {\it large} real number $\lambda$, 
this spectral weight will be strongly suppressed, exponentially in $\lambda$.
In this regime, the coefficient $c$ in the IR CFT Green's function behaves as (see 
eqn (D.22) of \citelow{Faulkner:2009wj} or 
eqn (5.18) of \citelow{Nabiltoappear})
\be c \propto e^{- \pi \lambda} \buildrel{\lambda \gg1 }\over{\to} 0 , \ee
and therefore the low-frequency spectral weight is
\be \rho_{{\rm osc}}(\omega, k ) \buildrel{\lambda \gg1 }\over{\to} \Imm 1 = 0 . \ee
}
This explanation is only a partial one because 
for given $g_m$, there is still an oscillatory region (at $k \propto g_m$), 
where there could in principle be gapless excitations.
\citelow{Edalati:2010ge} observes that even in that regime, 
where the IR CFT dimensions are $i$ times an order-one number, the spectral weight is suppressed.
An understanding of this effect must involve the behavior of the UV coefficients $a_{\pm}^{(0)}, b_{\pm}^{(0)}$
with $g_m$.

Despite this analytic partial understanding of the phenomenon observed in \citelow{Edalati:2010ww, Edalati:2010ge}, 
we retain some reservations about the interesting proposed connection to Mott physics.
In particular, the oscillatory region is playing a crucial role
in destroying the Fermi surface in this discussion.
Much about its interpretation remains mysterious, particularly in 
light of its implications\cite{Hartnoll:2009ns} for gravitational back-reaction.

Finally, we note that the same situation of no Fermi surfaces and large IR CFT dimension can also be reached 
by increasing $mL$, the mass of the spinor field in units of the $AdS$ radius, at fixed charge and no dipole coupling;
in this regime ($m > q e_d$) there is no oscillatory region.

\section*{Acknowledgments}

JM would like to thank Thomas Faulkner, Gary Horowitz, 
Nabil Iqbal, Hong Liu, Matthew Roberts, and David Vegh for 
collaboration on related matters.
We are grateful 
to Hong Liu, Mike Mulligan, and Brian Swingle for discussions
and to Raghu Mahajan and David Vegh for comments on the draft.
This work was supported in part by funds provided by the U.S. Department of Energy
(D.O.E.) under cooperative research agreement DE-FG0205ER41360,
and in part by the Alfred P. Sloan Foundation. 

\bibliographystyle{utphys}
\bibliography{randomcitations}

\end{document}